\documentclass[showpacs,preprintnumbers,amsmath,amssymb]{revtex4}

 \usepackage{epsfig}
\usepackage{amssymb}
\usepackage{graphicx}
\begin{document}
\title{ Interference of the Whispering Gallery  States of Antihydrogen}
\author{A. Yu. Voronin,  V.V. Nesvizhevsky, S. Reynaud}
\affiliation{ P.N. Lebedev Physical Institute, 53 Leninsky
prospect, Ru-117924 Moscow, Russia.
\\
Institut Laue-Langevin (ILL), 6 rue Jules Horowitz,
 F-38042, Grenoble, France.
\\
Laboratoire Kastler Brossel, Campus Jussieu, F-75252 Paris, France
}

\begin{abstract}

 We  study theoretically interference of the long-living quasistationary quantum states of antihydrogen atoms, localized near a concave material surface. Such states are an antimatter analog of the whispering gallery states of neutrons and matter atoms, and similar to the whispering gallery modes of sound and electro-magnetic waves. Quantum states of antihydrogen are formed by the combined effect of quantum reflection from  van der Waals/Casimir-Polder (vdW/CP) potential of the surface and the centrifugal potential. We point out a method for precision studies of quantum reflection of  antiatoms from vdW/CP potential; this method uses interference of the whispering gallery states of antihydrogen.

\end{abstract}

\maketitle

\section{Introduction}

 Localization of waves near a curved material surface, known as the whispering gallery (WG) effect \cite{Rayleigh,Ray1}, has been an object of growing interest during the last decade due to its multiple applications in optics \cite{Mie,Debye,Oraev,Vahal}, in atomic physics \cite{maser,Wallis,Aminoff,Mabichi,AtomTrapping,Bertram,CavityQED,Alton} and, since recently, in neutron physics  \cite{CentrProposal,Cubitt,Nature10,CentrNJP,CRP,NeutWG}. In the quantum mechanical description, the WG effect consists in evolution of a quantum object settled in long-living quasi-stationary states near a curved surface. Long life-times of such states and their small spatial size compared to the surface curvature radius, result in peculiar properties of the WG modes. In particular, neutrons or atoms, settled in the WG modes, spend relatively long time in the vicinity of a curved surface. Thus they might be sensitive to details of the atom(neutron)-surface potential and could serve as a probe for this interactions, in particular for establishing constraints for the strength of extra fundamental interactions with the spatial scale from a few nanometers to a micrometer \cite{PN,ShortRangeNeutron,Axion,Ant}, predicted in various extensions to the Standard Model. Weakly explored physics of the \emph{antiatom}-matter interaction is of particular interest. In this paper we study theoretically the interference between long-living WG states of antihydrogen ($\bar{H}$) atoms \cite{WGHbarStates} localized near a curved material surface in absence of external fields. We point out high sensitivity of an interference pattern to properties of the $\bar{H}$-matter interaction.

At a first glance the existence of any $\bar{H}$-matter long-living states seems improbable because of prompt annihilation of $\bar{H}$ in the bulk of a material wall. However, slow enough $\bar{H}$ atoms, impinging a material surface, are partially reflected due to the phenomenon of over-barrier quantum reflection \cite{LJ,Feinberg,Power,FJM} from the van der Waals/Casimir (vdW/CP) atom-wall potential \cite{CP,DLP,Shimizu,Pasquini,pasq06,Obrecht}.  The smaller is the normal incidence momentum $k$ of $\bar{H}$, the larger is the reflection coefficient. It follows from the Wigner law that the reflection coefficient tends to unity in the limit of zero $k$ \cite{voro05,voro05l}. This is why a $\bar{H}$ atom might be localized in quantum quasi-stationary states near a plane material surface \cite{GravStates} in the Earth's gravitational field. It "bounces" on a surface  the same way as neutrons bounce in gravitational quantum states, discovered recently \cite{nesv00,Nature1,nesv03,EPJC,Baes,UFN}. 
The characteristic lifetime of $\bar{H}$ states above a plane conducting surface in the Earth's gravitational field  is long enough ($\tau\simeq 0.1$ s) and could be even significantly increased by choosing a proper material.  

In the related similar phenomenon of  localization of $\bar{H}$  atoms moving in the vicinity of a curved  surface, considered in this paper, corresponding quasi-stationary quantum states are formed by  superposition of the effective centrifugal potential and the quantum reflection. WG states are the states with high angular momentum, so that the  kinetic energy of tangential motion (parallel to the curved surface) is close to the total energy. This means that $\bar{H}$  radial motion normal to the material wall is slow, which  provides large probability of quantum reflection and thus long lifetime of the quasi-stationary states.
  For a certain range of  the tangential velocities $v$ of $\bar{H}$   the effective centrifugal potential can be  approximated quite accurately by a linear potential in the vicinity of the curved surface. In the problem of interest we deal with the effective centrifugal potential $(mv^2/R) x$ (where $m$ is the inertial mass of $\bar{H}$, $x$ is the distance from the surface),  instead of the gravitational potential $Mgx$ (where $M$ is the gravitational mass of $\bar{H}$ and $g$ is the free fall acceleration).  Thus the WG states appear complementary to gravitational states in testing WEP in the quantum domain \cite{QEPL,QEPOnofrio}. Simplicity to tune the effective centrifugal acceleration by changing the (anti)atom beam velocity $v$ is an important advantage of studying WG states.

In a long standing problem of experimental testing of antimatter gravitational properties \cite{PositronFall, PositronFallN, AntiprotonFall,cesa05,gabr10, Aegis1} the phenomenon of existence of  quasi-stationary states of $\bar{H}$,  localized near material surface, could become a novel promising tool \cite{Yam}. Indeed,  clearly visible  antiatom-wall annihilation signal might make possible observation of  quantum interference effects for such states, which should be sensitive to antimatter gravitational properties \cite{GravStates}. Such studies would require, as an important ingredient, the knowledge  of the antiatom-material wall scattering length, which characterizes the antiatom-matter interaction in the considered low energy limit. The interferometry of  WG states of $\bar{H}$ atoms,  could be an independent and  precision method to measure these antiatom-matter scattering lengths for different materials.

  The paper is organized as follows. In section \ref{section1}, we write down the main equations describing WG states, and estimate the spatial and energy scales of the problem. In section \ref{section2}, we study theoretically the transmission of $\bar{H}$ throw a slit between two coaxial cylindrical surfaces and point out a precision method to study the quantum reflection. In section \ref{section3}, we study theoretically the interference between antihydrogen WG states
and demonstrate potentially high sensitivity of the interference pattern to parameters of the system. In Appendix  we present details of derivation expressions for the WG states energies of $\bar{H}$ between two coaxial cylinders.

\section{Antihydrogen WG states} \label{section1}

In the following we will be interested in a certain state of motion of an $\bar{H}$ atom along the surface of a curved material mirror, during which $\bar{H}$ atom undergoes multiple \textit{grazing} collisions with a material surface. The  interaction of $\bar{H}$ atom with the material wall could be described by vdW/CP potential. Such a potential has the asymptotic $C_4/R^4$,  where $R$ is the distance from the wall and $C_4$ is the Casimir-Polder coefficient \cite{CP}, which depends on the wall material properties alone and stands for the retardation effects in the (anti)atom-wall interaction. This potential  can be characterized by the  spatial scale $l_{CP}$ \cite{voro05}:
\begin{eqnarray}
l_{CP}=\sqrt{2 m C_4}\mbox{, }
l_{CP}=0.027 \mbox{ } \mu m,
\end{eqnarray}
$m$ is the mass of an $\bar{H}$ atom.
The prompt annihilation of $\bar{H}$ in the bulk of the wall can be described by means of a full absorbtion boundary condition \cite{voro05,voro05l}. According to the Wigner law, the reflection coefficient $r$ for  the (anti)atom-wall system in the low energy limit is determined by the imaginary part of the scattering length $a_{CP}$ of such an interaction:
\begin{equation}\label{Refl}
     r\simeq 1-4 k |\mathop{\rm Im} a_{CP}|/\hbar,
      \end{equation}
      $k$ is the  transversal momentum of $\bar{H}$.

The values of $a_{CP}$ were calculated in  \cite{voro05} for the case of perfectly conducting surface:
 \begin{eqnarray}
a_{CP}=-(0.10+i1.05)l_{CP} \mbox{, }
a_{CP}=-0.0027-i0.0287 \mu m.
\end{eqnarray}
The   value $a_{CP}$  characterizes  completely the absorptive(reflective) properties of the material wall in case of low energy $\bar{H}$-wall collisions.

As follows from eq.(\ref{Refl}) , if the  transverse momentum of $\bar{H}$ is small, so that the condition $k |\mathop{\rm Im} a_{CP}|/\hbar\ll 1$ is valid, $\bar{H}$ atoms are reflected efficiently from the material wall. This reflection explains  coexistence of  slow $\bar{H}$ atoms and material surroundings in a form of quasi-stationary states. In the following we will apply these arguments to WG states of $\bar{H}$ atoms.

(Anti)atom dynamics in a cylindrical wave-guide with the radius $R$ obeys the following Schr\"{o}dinger equation in the cylindrical coordinates:

\begin{equation}
\label{SchrEq1}
\left[-\frac{\hbar^2}{2m}\left(\frac{\partial^2}{\partial\rho^2}\right)-
\frac{\hbar^2}{2m\rho^2}\left(\frac{\partial^2}{\partial\varphi^2}+\frac{1}{4}\right)+V_{CP}(|R-\rho|)-\frac{p^2}{2m}\right]\Phi(\rho,\varphi)=0.
\end{equation}
Here $\Phi(\rho,\varphi)$ is the  wavefunction of $\bar{H}$,   $\rho$ is the radial distance  measured from the cylinder axis,
 $\varphi$ is the angle,  $V_{CP}(|R-\rho|)$ is the $\bar{H}$-wall vdW/CP interaction potential \cite{voro05l,DMB} and $p$ is the  momentum of $\bar{H}$. We omit the trivial dependence on $z$-coordinate along the cylinder axis in this equation. The above equation meets the following boundary conditions: regularity at $\rho=0$, uniqueness under the $2\pi$-shift in the angle $\varphi\rightarrow\varphi+2\pi$, and full absorption (annihilation) in the wall bulk at distances $\rho\geq R$ \cite{voro05l}.

The wave-function is decomposed using the angular momentum eigenfunctions basis as follows:
\begin{equation}\label{Decomp}
 \Phi(\rho,\varphi)=\sum_{\mu=-\infty}^{\mu=+\infty}\chi_{\mu}(\rho)\exp(i \mu \varphi),
\end{equation}
with $\chi_{\mu}(\rho)$ the radial motion wave-functions.
The solutions of interest are regular at the zero distance $\rho=0$.

One can show that WG states are  the states with large angular momentum \cite{WGHbarStates} $\mu$ such that $\hbar^2 \mu^2/R^2\approx p^2$ \cite{CentrNJP}. Indeed, in this case the radial motion is slow that is the necessary condition for the efficient  quantum reflection of $\bar{H}$ from the surface.
The typical values of angular momentum $\mu$ for practically important (see below) values of the tangential velocity of $\bar{H}$ $v=1$ m/s, and the cylinder radius  $R=0.1$ m is $\mu\simeq 1.6\times10^6$. This means that $\bar{H}$ motion in WG states of interest along the angular coordinate $\varphi$ is semiclassical. This fact enables us to consider a motion parallel to the surface as classical and introduce  temporal dependence of the wave-function $\Phi (\rho,t)$  assuming that $\varphi=v t/R$. In the following we will use such a temporal treatment.

 We expand the expression for the centrifugal potential in the vicinity of the distance $\rho=R$ introducing the deviation $x$ from the cylinder surface $x=\rho-R$. In the first order of the small ratio $x/R$, we get the following Schr\"{o}dinger equation:
\begin{equation}\label{Rexp}
\left[-\frac{\hbar^2}{2m}\frac{\partial^2}{\partial
x^2}+V_{CP}(-x)+\hbar^2\frac{\mu^2-1/4}{2m
R^2}\left(1-\frac{2x}{R}\right)-\frac{p^2}{2m}\right]\chi_{\mu}(x)=0.
\end{equation}

Introducing the radial motion energy $\varepsilon_{\mu}$:
$$\varepsilon_{\mu}=\frac{p^2}{2m}-\hbar^2\frac{\mu^2-1/4}{2mR^2}\simeq\frac{(pR)^2-\hbar^2\mu^2}{2mR^2},$$ 
and the tangential velocity $v_{\mu}$, so that $\mu=mv_{\mu}R/\hbar$, we translate the Schr\"{o}dinger equation into the form:
\begin{equation}\label{RexpLinear}
\left[-\frac{\hbar^2}{2m}\frac{\partial^2}{\partial
x^2}+V_{CP}(-x)-\frac{mv_{\mu}^2}{R}x-\varepsilon_{\mu}\right]\chi_{\mu}(x)=0.
\end{equation}
 In the following we will omit the subscript $\mu$, assuming that the results are obtained for a fixed $\mu$ value.

Eq. (\ref{RexpLinear}) defines the spatial  $l_0$ and energy $\varepsilon_0$ scales, characteristic for the effective linear potential $m (v^2/R) x$:
\begin{eqnarray}\label{L0}
l_0&=&\sqrt[3]{\frac{\hbar^2R}{2m^2v^2}}, \mbox { and} \\
\varepsilon_0&=&\sqrt[3]{\frac{\hbar^2 m v^4}{2R^2}}.\label{E0}
\end{eqnarray}

Efficient quantum reflection of $\bar{H}$ occurs only in case of hierarchy of the spatial scales of  vdW/CP potential $l_{CP}$ and the linear (centrifugal) potential spatial scale $l_0$ :
\begin{equation}\label{Hierarchy}
l_0\gg l_{CP}.
\end{equation}

This condition means, in particular, that the effect of  vdW/CP potential on the  motion of $\bar{H}$ in a linear attractive potential can be described using a modified boundary condition at $x=0$. Such a  boundary condition at $x=0$ ensures the matching of logarithmic derivative of the WG state wave-function with the solution inside the vdW/CP potential and has the form \cite{voro05l}:
\begin{equation}\label{Boundary}
\frac{\chi( 0)}{\chi'( 0)}=-a_{CP}.
\end{equation}
  So far the reflecting (absorbing) properties of the material wall appear in the above formalism only via the constant $a_{CP}$.
  If the condition Eq.(\ref{Hierarchy}) is valid the problem of interest is  equivalent to the  problem of $\bar{H}$ motion in a superposition of the linear \textit{gravitational} potential and the CP/vdW antiatom-wall interaction potential studied in \cite{GravStates}. In particular, for the parameters  $R=0.1$ m, $v=0.99$ m/s, the effective centrifugal acceleration coincides with the free fall acceleration $v^2/R=g$. Then the corresponding whispering gallery spatial scale is $l_0=5.87$ $\mu$m, while the energy scale is $\varepsilon_0=0.60$ peV.
For the lowest states, the  eigen-functions are well-known  Airy functions \cite{abra72} with a complex shifted argument:
\begin{equation}\label{EigenFunct}
    \chi_n(x)\sim\mathop{\rm Ai}(x/l_0-\lambda_n),
\end{equation}
where $\lambda_n$ are the modified eigen-values of a quantum bouncer, which are the solutions of the following equation:
\begin{equation}\label{EigenValues}
\mathop{\rm Ai}(-\lambda_n+a_{CP}/l_0)=0.
\end{equation}
The corresponding complex energies of quasi-stationary states are:
\begin{equation}\label{Energies}
\varepsilon_n=\varepsilon_0 \lambda_n^0+ \frac{mv^2}{R}a_{CP},
\end{equation}
where $\lambda_n^0$ are the eigen-values of the standard quantum bouncer (i.e. a quantum particle in a linear potential $V(x)\sim x$ superimposed with a totally reflecting mirror at $x=0$ ), which obey the following equation:
\begin{equation}\label{EigenValue0}
 \mathop{\rm Ai}(-\lambda_n^0)=0.
\end{equation}
The width of\textit{ all} such states is equal:
\begin{equation}\label{Gamma}
    \Gamma/2=\frac{mv^2}{R}|\mathop{\rm Im}a_{CP}|.
\end{equation}

In particular,  the corresponding lifetime $\tau=\hbar/\Gamma$ of quasi-stationary states equals $0.1$ s for the velocity $v=0.99$ m/s and mirror radius $R=0.10$ m for an ideally conducting surface, and it is $0.2$ s for a silica surface \cite{Reynaud}. The  angular position of the $\bar{H}$ atom is shifted by:
\begin{equation}\label{DeltaPhi}
 \Delta \varphi=\frac{\tau v}{R}=\frac{\hbar}{2 mv |\mathop{\rm Im}a_{CP}|}
\end{equation}
during this lifetime. $\Delta \varphi=1.136$ rad for the mentioned above parameters. For a silica surface this value is twice larger.

Properties of whispering gallery states of $\bar{H}$ near a cylindrical surface are similar to the gravitational states of $\bar{H}$ near a plane surface. However, only inertial mass of $\bar{H}$ is involved in the case of whispering gallery states; also the effective acceleration $v^2/R$ is a tunable parameter.

\section{An integral method to measure  the $\bar{H}$-wall reflection properties }\label{section2}

As already mentioned  the  value $a_{CP}$  characterizes absorptive(reflective) properties of a material wall in the limit of low energy $\bar{H}$ atom scattering. In this section we present a theoretical formalism describing transition of $\bar{H}$ atoms through a slit between two coaxial cylindrical surfaces and propose to  use this configuration for precision measurement of $|\mathop{\rm Im}a_{CP}|$, 
 thus getting access to studies of antiatom-material surface interactions.

\begin{figure}[h!]
 \centering
 \includegraphics[width=115mm]{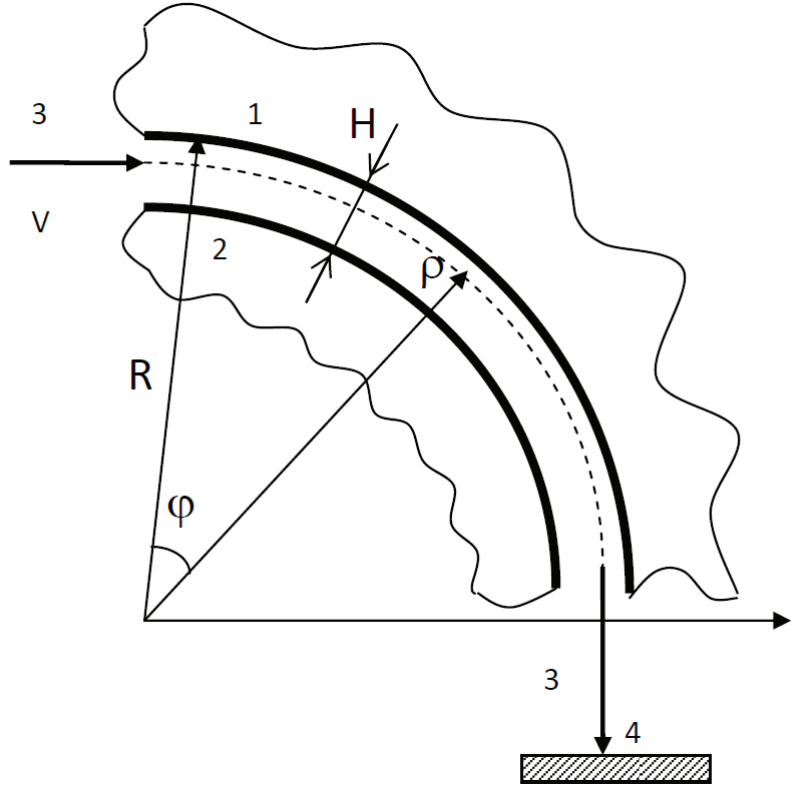}
\caption{A sketch of the experimental setup. 1 indicates outer cylindric mirror, 2 indicates inner cylindrical mirror, 3 indicates the   momentum of $\bar{H}$ atoms, parallel to the mirror surface, 4 indicates the detector of $\bar{H}$ at the exit of the mirrors assembly. The count rate of $\bar{H}$ atoms at the exit of the mirror assembly is measured as a function of velocity of $\bar{H}$ atoms.  }\label{ExperSetup}
\end{figure}

A sketch of a posible experiment is shown in Fig.\ref{ExperSetup}. This scheme combines two approaches used previously in experiments on gravitational and whispering-gallery states of slow neutrons: an assembly of two mirrors (one of which plays a role of a scatterer/absorber) as in \cite{Nature1} and curved mirror  surfaces (instead of  flat ones) as in \cite{Nature10}. The concept of the measurement would be in this case analogous to that, used in \cite{Nature1} for studying gravitational states of neutrons, namely to measure the count rate of $\bar{H}$ atoms penetrating through the slit between mirrors as a function of the ratio of the spatial quantum state size and the slit size. Only  states, with the spatial size  smaller than the slit size could easily penetrate through the slit, other states would be effectively absorbed.

An advantage of such an approach compared to measurements of single-reflection effects is much longer antiatom-surface interaction time provided antiatoms are settled in whispering gallery states. On the other hand, the  antiatom tangential velocity turns to be an important tunable parameter in such a measurement \cite{Nature10,CentrNJP,Cubitt}.  

    In the previous section we found that if the tangential velocity $v$ is \textit{fixed}, then evolution of $\bar{H}$, settled in a superposition of whispering gallery states, is described by eq.(\ref{RexpLinear}). An equivalent problem of ultra-cold neutron motion in the Earth's gravitational field has been treated in \cite{NVP,MNPRA,MeyNesv2,Esc}. The corresponding equations for $\bar{H}$ evolution in whispering gallery states  between two coaxial cylindrical segments  are:
 \begin{equation}\label{2cylinders}
 \left\{\begin{array}{clll}\left[ -\frac{\hbar^2 \partial ^{2}}{2m\partial x^{2}}+\frac{mv^2}{R}x-\widetilde{\varepsilon}\right] \chi_{\mu}
(x)=0\\
\frac{\chi( 0)}{\chi'( 0)}=-a_{CP}\\
\frac{\chi( H)}{\chi'(H)}=-a_{CP}.
\end{array}
\right.
\end{equation}
 We apply here  two boundary conditions of the type eq.(\ref{Boundary}) at $x=0$ and $x=H$ to describe partial reflection of $\bar{H}$ atoms from the exterior ($x=0$) and interior cylinders ($x=H$).  This equation system determines  complex eigen-energies $\widetilde{\varepsilon}_n$ of quasi-stationary states of $\bar{H}$ localized between two cylinders. We derive the expressions for these eigen-energies and discuss their properties in Appendix.


An important property of the corresponding eigen-energies $\widetilde{\varepsilon}_n$
is that the state width is  changing sharply as a function of the quantum number $n$. For the quantum numbers $n$ smaller than $N$, such that $H\simeq\lambda_N^0 l_0$, the eigen-values $\widetilde{\varepsilon}_n$ are close to those of WG states, given by eq.(\ref{Energies}), $ \widetilde{\varepsilon}_n\rightarrow \varepsilon_n$, $n\leq N$. Indeed, in such a case the centrifugal barrier prevents $\bar{H}$ in lower states  $n\leq N$ to penetrate into the second cylinder, so the $\bar{H}$ atoms are mainly scattered by the exterior cylinder.

\begin{figure}[h!]
 \centering
 \includegraphics[width=115mm]{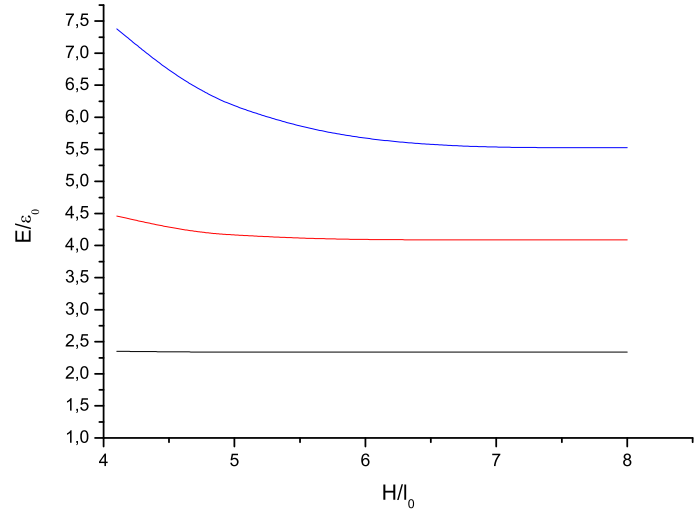}
\caption{The energies of the first three whispering gallery states as a function of the distance between the cylinder surfaces $H$, in dimensionless units.}\label{energy}
\end{figure}

\begin{figure}[h!]
 \centering
 \includegraphics[width=115mm]{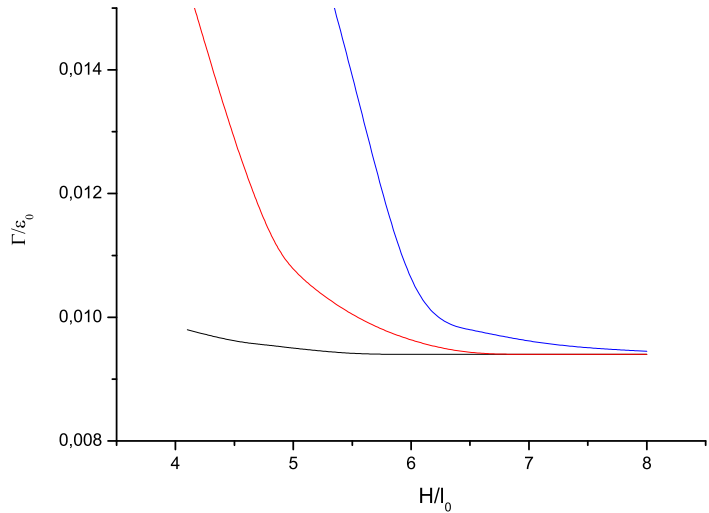}
\caption{The width of the first three whispering gallery states as a function of the distance between the cylinder surfaces $H$, in dimensionless units.}\label{width}
\end{figure}

\begin{figure}[h!]
 \centering
 \includegraphics[width=115mm]{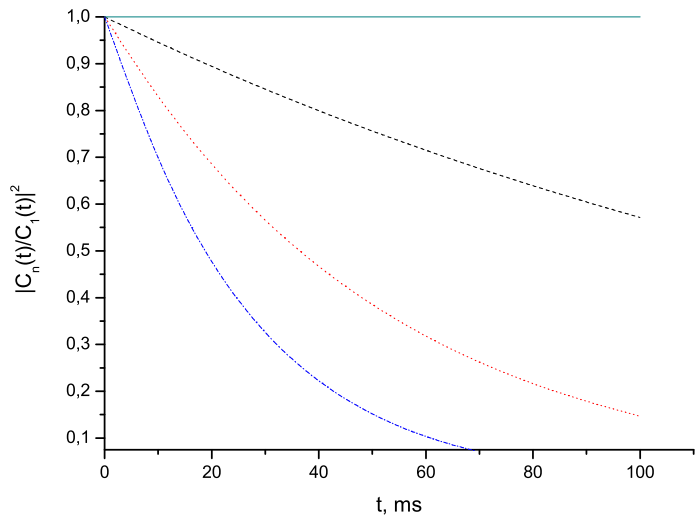}
\caption{Relative contributions  of the four lowest whispering gallery states to the $\bar{H}$ density as a function of time. Solid line corresponds to the relative contribution of  first state, dashed line stands for second state, dotted line indicates third state, and dash-dotted line shows forth state. }\label{decay}
\end{figure}

For  $n>N$ $\bar{H}$ atoms penetrate easily to the interior cylinder and are settled in states formed by partial reflection from both cylinder surfaces. The width of such states is much larger. We show in Figs.(\ref{energy},\ref{width}) the energy and width of the first three states as a function of the distance $H$ between the cylinder surfaces.
One can see a sharp increase in the width of the states if $H<\lambda_n^0 l_0$. This sharp dependence of the states lifetime could be used to prepare a superposition of a given number of whispering gallery states, as far as the states with higher $n$ decay much faster. The squares of relative amplitudes $C_n(t)$  of four lowest states  are shown in Fig.(\ref{decay}) as a function of time.
The number of states $N$, which  contribute significantly to the flux through the waveguide is thus given by the number of states with the spatial size smaller than $H$. These states have quantum number $n$ such that $n<N, H\approx\lambda_N^0 l_0$.
    One can show that for $N\gg1$ it could be approximated by the following semiclassical expression:
    \begin{equation}\label{Nstates}
     N \approx \frac{(2H)^{3/2}m v}{3\hbar \pi \sqrt{R}}
     \end{equation}
    Thus in the above limit of large $N$, it is proportional to the tangential velocity of antiatoms.

    The coefficient $F$ of transmission through the wave-guide, studied above, is given by:
    \begin{equation} \label{Ftrans}
    F\approx \sum_{i=1}^{\infty} |C_{i}(T)|^2 \approx \sum_{i=1}^{\infty} |C_{i}(0)|^2\exp{(-\Gamma_i T/\hbar)}.
    \end{equation}
    Here  $T=L/v$ is the classical time of flight along the curved path with the length $L$. In the above expression we neglected small interference terms, which result from nonorthogonality of quasi-stationary whispering gallery states (see discussion in \cite{NVP}). Taking into account eq.(\ref{Gamma}) for the width of whispering gallery states (equal for all states) and limiting the sum with contribution of states with $n\leq N$, we get:
     \begin{equation}\label{Fapr}
     F\approx \frac{(2H)^{3/2}m }{3\hbar \pi \sqrt{R}}  v \exp{\left[2 m v \mathop{\rm Im}a_{CP} L/(\hbar R)\right]}
     \end{equation}
     for the transmission coefficient. An important consequence of the above expression, which is asymptotically valid for large $N$, is exponential dependence of the transmission coefficient as a function of a factor $\mathop{\rm Im}a_{CP}$.
      The sharp dependence of $F$ on $ \mathop{\rm Im}a_{CP}$  follows from long antiatom-surface interaction time (multiple reflections) in the whispering gallery mode.

\begin{figure}[h!]
 \centering
 \includegraphics[width=115mm]{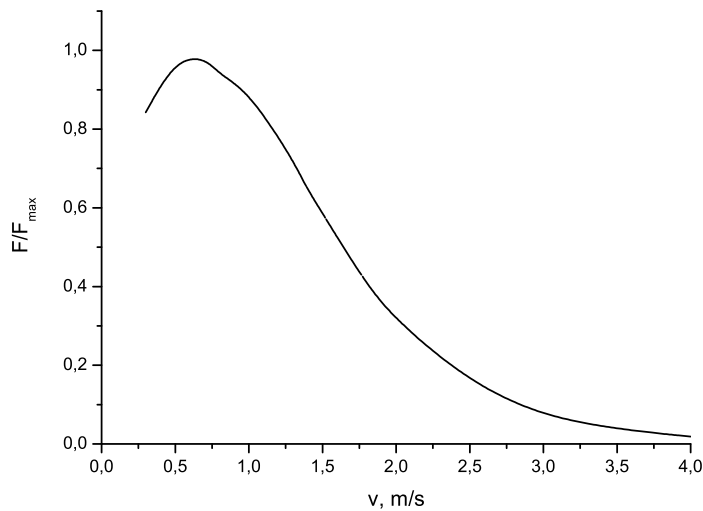}
\caption{The transmission coefficient is shown as a function of the tangential velocity v. The waveguide consists of two coaxial cylinder with the external radius $R=0.1$ m, the distance between the cylinders $H=52$ $\mu$m, and the path length $L=0.15$ m.}\label{TransFig}
\end{figure}

     We show in Fig.\ref{TransFig} the coefficient of transmission through the waveguide, consisting of two coaxial cylinders with the external radius $R=0.1$ m, the distance between the cylinders $H=52$ $\mu$m, and the curved path length $L=0.15$ m, as a function of the $\bar{H}$ tangential velocity $v$. We obtained the corresponding values  by numerical solution of Eq. (\ref{2cylinders}).

The choice of suggested parameters  is conditioned by the following arguments. The  velocity $v$ of $\bar{H}$ atoms is chosen approximately equal to the velocities expected in experiments on production of ultracold $\bar{H}$ \cite{Yam}. The distance between the cylinders allows selection of only a few quantum states. Their characteristic size $l_0$ is given by eq.(\ref{L0}) and depends on the velocity $v$ and the cylinder mirror radius. It is chosen to be equal to the analogous  gravitational length scale in experiments with ultracold neutrons \cite{Nature1}. Much smaller values of $l_0$ would not allow precise mechanical adjustment, while significantly larger values would correspond to too low energies of the WG states and larger systematic effects accordingly. Thus the choice of parameters of the  mirror assembly and atom velocity corresponds to those parameters, which are likely to be used in experiments with $\bar{H}$.

\section{An interference method to study whispering gallery states}\label{section3}
Studies of interference patterns is known to be a powerful method for precision measurement of  properties of quantum systems. We suggest to study the interference of whispering gallery states in order to get information about properties of variety of antiatom-matter interactions. A construction of a wave-packet, in which only a few quantum states are superimposed, is a necessary condition for clearly observing interference. We use the waveguide consisting of two co-axial cylinders which, as we have shown, enables us to construct a wave-packet of a few states. The number of such states is determined by the distance $H$ between cylinders and the tangential velocity $v$, according to eq.(\ref{Nstates}).

One can show \cite{GravStates} that the decay character of $\bar{H}$ states near a material surface results in nonzero current through the surface. Indeed, the boundary condition at the surface eq. (\ref{Boundary}) includes the \textit{complex} scattering length $a_{CP}$, which guarantees the non-zero current, calculated at the surface position $x=0$:
\begin{equation}\label{j}
j(0,t)=\frac{i\hbar}{2m}\left(\Phi(0,t) \frac{d \Phi^*(0,t)}{dx}-\Phi^*(0,t) \frac{d \Phi(0,t)}{dx}\right).
\end{equation}
Such a current determines the disappearance rate $dF(t)/dt$ of $\bar{H}$. An interesting property of the rate of disappearance of a \textit{superposition of several} quantum states is the effect of temporal beatings, determined by differences in the states energies. These beatings manifest in observable change of the number of annihilation events as a function of time; they allow for measurement of the frequencies of transitions between energy levels. Using the properties of Airy functions one can show \cite{GravStates} that

\begin{equation}\label{Nt}
 \frac{dF(t)}{dt}=-\frac{\Gamma}{\hbar} \exp(-\frac{\Gamma}{\hbar} t) \left(\sum_{i}^n |C_i|^2+2\mathop{\rm
 Re}\sum_{i>j}^n\sum_{j}^n C_{j}^*C_{i}\exp(-i(\varepsilon_i-\varepsilon_j)\frac{t}\hbar)\right).
 \end{equation}

Here $C_i$ is the amplitude of the corresponding quantum state in the superposition of interest.
 For the case of interference of three lowest equally populated states, studied below, the expression for the disappearance rate turns to be:
 \begin{equation}\label{3st}
 \frac{dF_{123}(t)}{dt}\approx-\frac{2}{3}\frac{\Gamma}{\hbar} \exp(-\frac{\Gamma}{\hbar}
 t)\left(\frac{3}{2}+\cos(\omega_{12}t)
 +\cos(\omega_{23}t)+\cos((\omega_{12}+\omega_{23})t)\right),
 \end{equation}
where $\omega_{ij}=(\varepsilon_j-\varepsilon_i)/\hbar$.
The large-scale modulation of the interference beatings is given by the period of coherence of $\cos(\omega_{12}t)$ and $\cos(\omega_{23}t)$ terms:
\begin{equation}\label{Trev}
T_{r}=\frac{2\pi}{\omega_{12}-\omega_{23}}.
\end{equation}
As  partial frequencies $\omega_{12}$ and $\omega_{23}$ are close to each other, the period $T_r$ is  large compared to the characteristic period of beatings. The value of $T_r$ is sensitive to  small variations of partial frequencies, produced by any additional interaction in the system. It is also sensitive to the variation of the beating time scale $\hbar/\varepsilon_0$, induced by the change of the tangential velocity $v$.

\begin{figure}[h!]
 \centering
 \includegraphics[width=115mm]{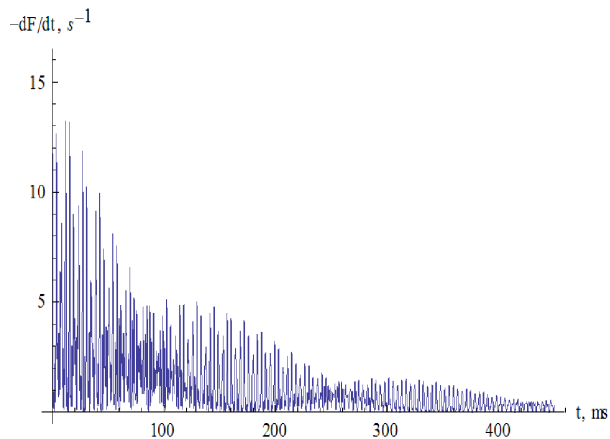}
\caption{The $\bar{H}$ disappearance rate as a function of time for a superposition of several whispering gallery states. $R=0.1$ m, $H=32$ $\mu$m $v=1.00$ m/s.}\label{interference1}
\end{figure}

\begin{figure}[h!]
 \centering
 \includegraphics[width=115mm]{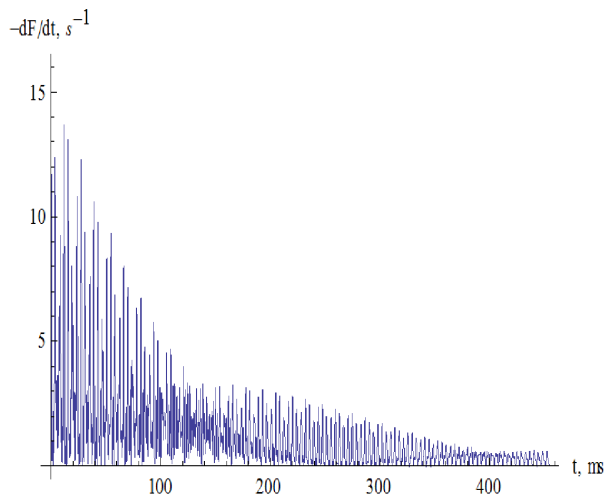}
\caption{The $\bar{H}$ disappearance rate is shown as a function of time for a superposition of several whispering gallery states. $R=0.1$ m, $H=32$ $\mu$m $v=0.99$ m/s.}\label{interference2}
\end{figure}

 To demonstrate sensitivity of an interference pattern to parameters of the quantum system we show in Fig.(\ref{interference1}) and in Fig.(\ref{interference2}) the disappearance rate as a function of time for two different tangential velocities of $\bar{H}$ equal to $1$ $m/s$ and $0.99$ $m/s$. The difference in the large-scale modulation character of the interference pattern is clearly seen.


 In the studied above cases  $T_{r}=0.170$ s for the tangential velocity  $v=1.00$ $ m/s$, and $T_r=0.265$ s for the tangential velocity $v=0.99$ $m/s$ and the difference in these values determines the visible difference in interference patterns in Fig.(\ref{interference1}) and Fig.(\ref{interference2}).

 We estimate that no major systematic effects would appear in experiments, modeled in Fig.\ref{TransFig},\ref{interference1} and \ref{interference2} provided the quality of mirrors and the experimental setup are equivalent to those actually used in already performed analogous experiments with ultracold neutrons \cite{Nature1, Nature10}.

\section {Conclusion}

We have studied theoretically interference between WG states of $\bar{H}$ atoms, moving near a cylindrical material surface. Such states are an antimatter wave analog of the whispering gallery states of neutrons and matter atoms. The localization of $\bar{H}$ atoms near a surface of a curved mirror is due to the combined effect of quantum reflection and centrifugal potential. We show that the rate of $\bar{H}$ annihilation on the material surface has to exhibit temporal beatings, which period is determined by difference in the whispering gallery states energies. This fact would allow for precision measurements of the $\bar{H}$-matter interaction properties, using interferometric methods. The benefit of such an approach is due to long time of interaction with surface for anti-atoms settled in whispering gallery states. We have pointed out a method to study the $\bar{H}$-matter quantum reflection  by measuring the coefficient of transmission through a curved wave-guide. Thus obtained scattering properties of material wall is an important ingredient in precision studies of gravitational properties of $\bar{H}$ atoms, localized in gravitational quantum states above the material surface \cite{Yam}. We have proposed a way to shape a superposition of a few such states and to observe their interference in the temporal behavior of the annihilation rate. 
The whispering gallery states of $\bar{H}$ could be a promising tool for precision measurements of the $\bar{H}$-surface interactions, gravitational properties and WEP tests, guiding and trapping antiatoms.

\section {Appendix} \label{Appendix}
In this Appendix we derive the asymptotic properties of the eigen-energies of the WG states between two coaxial cylinders, determined in eq. (\ref{2cylinders}).
The solution of Eq.(\ref{2cylinders}) is a linear combination of Airy functions:
\begin{equation}\label{LinearCombination}
   \chi_n(x)\sim\mathop{\rm Ai}(x/l_0-\beta_n)-S\mathop{\rm Bi}(x/l_0-\beta_n).
\end{equation}
We introduced here the dimensionless eigenvalues $\beta_n=\widetilde{\varepsilon}_n/\varepsilon_0$.

Boundary conditions at $x=0$ and $x=H$ imply the following condition for the eigenvalues $\beta_n$:
\begin{equation}\label{EigenValues2}
\frac{\mathop{\rm Ai}(-\beta_n+a_{CP}/l_0)}{\mathop{\rm Bi}(-\beta_n+a_{CP}/l_0)}=\frac{\mathop{\rm Ai}(H/l_0-\beta_n+a_{CP}/l_0)-a_{CP}/l_0 \mathop{\rm Ai'}(H/l_0-\beta_n+a_{CP}/l_0)}{\mathop{\rm Bi}(H/l_0-\beta_n+a_{CP}/l_0)-a_{CP}/l_0 \mathop{\rm Bi'}(H/l_0-\beta_n+a_{CP}/l_0)}.
\end{equation}

It is interesting to get the asymptotic expression for the eigen-values $\beta_n$ as a function of distance $H$ between cylinders, when $H> |\beta_n|l_0$. This condition provides that the interior cylinder surface is located within the classically forbidden region for a given state $\beta_n$.
Using the asymptotic behavior of the Airy functions  $x \gg \beta_n$ :
  \begin{eqnarray}\label{AssAi}
 \mathop{\rm Ai}(x-\beta_n)&\sim&\frac{1}{\sqrt[4]{x-\beta_n}} \exp(-2/3 (x-\beta_n)^{3/2})\\
 \mathop{\rm Bi}(x-\beta_n)&\sim&\frac{1}{\sqrt[4]{x-\beta_n}} \exp(2/3 (x-\beta_n)^{3/2}),
 \end{eqnarray}

we get the following approximative equation for the eigen-values $\beta_n$:
\begin{equation}\label{High}
 \frac{\mathop{\rm Ai}(-\beta_n+a_{CP}/l_0)}{\mathop{\rm Bi}(-\beta_n+a_{CP}/l_0)}\simeq\frac{1}{2}Exp\left(-\frac{4}{3}(H/l_0-\beta_n)^{3/2}\right) \left(1+2\frac{a_{CP}}{l_0}\sqrt{H/l_0-\beta_n}\right).
\end{equation}
It follows from Eq.(\ref{High}) that in the limit $H\gg|\beta_n| l_0$ :
\begin{equation}\label{betaHigh}
  \mathop{\rm Ai}(-\beta_n+a_{CP}/l_0)\rightarrow 0,
\end{equation}
which is equivalent to the equation for the eigen-values $\lambda_n$ of WG states, reflected from one cylinder surface:
\begin{equation}
  \mathop{\rm Ai}(-\lambda_n+a_{CP}/l_0)= 0,
\end{equation}

 hence in the limit $H\gg|\lambda_n| l_0$ the eigen-values $\beta_n$ tend to $\beta_n\rightarrow \lambda_n$.

Using the asymptotic equation for the eigen-values $\beta_n$, given in eq.(\ref{High}), it is possible to correct  the eigen-values $\lambda_n$ of WG states  due to the presence of the interior cylinder in the classically forbidden region:
\begin{equation}\label{Corrlamb}
 \beta_n=\lambda_n-\frac{\mathop{\rm Bi}(-\lambda_n)}{2\mathop{\rm
  Ai'}(-\lambda_n)}\exp \left[-\frac{4}{3}(
H/l_0-\lambda_n)^{3/2}\right]\left(2\sqrt{H/l_0-\lambda_n}\frac{a_{CP}}{l_0}+1\right)
 \end{equation}
 The width of the $n$-th state due to penetration
 under the gravitational barrier and absorption in the second cylinder turns out to be:
 \begin{equation}\label{Gfl}
 \Gamma_n\simeq 2\frac{|\mathop{\rm
 Im}a+{CP}|}{l_0}\varepsilon_0\left(1+\sqrt{(H/l_0-\lambda_n)/\lambda_n}\exp
 \left[-\frac{4}{3}(H/l_0-\lambda_n)^{3/2}\right]\right)
 \end{equation}
  One can see that the width of a given state of a $\bar{H}$ atom between two cylinders increases steeply when the state spatial size ( which can be estimated as a position of the classical turning point)  approaches the position of the interior cylinder surface $\lambda_n l_0 \rightarrow H$.

  In the opposite limit $H\ll \lambda_n l_0 $ the centrifugal potential can be neglected and one has a case of a "box-like" state, which energy level is determined by the following equation system:
   \begin{equation}\label{box-like}
 \left\{\begin{array}{clll}\left[ \frac{\hbar^2 \partial ^{2}}{2m\partial x^{2}}+ \frac{\kappa_n^2}{2m}\right] \chi_{\mu}
(x)=0\\
\frac{\chi( 0)}{\chi'( 0)}=-a_{CP}\\
\frac{\chi( H)}{\chi'(H)}=-a_{CP}.
\end{array}
\right.
\end{equation}
The above equation is satisfied for the values$\kappa_n$:
\begin{equation}
 \kappa_n=\frac{\pi n}{H+2 a_{CP}}
 \end{equation}

The corresponding eigen-energies $\widetilde{\varepsilon}_n=\hbar^2\kappa^2/(2m)$ in the limit $H\ll \lambda_n l_0 $ turn to be:
\begin{equation}
\widetilde{\varepsilon}_n\simeq \frac{\hbar^2\pi^2 n^2}{2m H^2}\left(1-4 a_{CP}/H\right)
\end{equation}

\bibliographystyle{unsrt}
\bibliography{hbarclock}

\end{document}